%% file: main.tex
\documentclass{INTERSPEECH2023}

\usepackage{amsmath,graphicx}
\usepackage{multirow,array,enumitem,adjustbox}
\usepackage{booktabs,threeparttable,makecell}
\usepackage{xcolor}
\usepackage{cleveref}
\usepackage[linesnumbered,ruled,vlined]{algorithm2e}

\SetCommentSty{mycommfont}

\interspeechcameraready

\title{Attention-based Encoder-Decoder Network for End-to-End Neural Speaker Diarization with Target Speaker Attractor}
\name{Zhengyang Chen$^1$, Bing Han$^1$, Shuai Wang$^2$, Yanmin Qian$^{1\dagger}$ \thanks{$^\dagger$Corresponding Author} 
}

\address{
  $^1$MoE Key Lab of Artificial Intelligence, AI Institute \\
  X-LANCE Lab, Department of Computer Science and Engineering, Shanghai Jiao Tong University \\
  $^2$Shenzhen Research Institute of Big Data, Shenzhen, China
  }
\email{\{zhengyang.chen, hanbing97, yanminqian\}@sjtu.edu.cn, wangshuai@cuhk.edu.cn}

\begin{document}

\maketitle
 
\begin{abstract}
This paper proposes a novel Attention-based Encoder-Decoder network for End-to-End Neural speaker Diarization (AED-EEND). 
In AED-EEND system, we incorporate the target speaker enrollment information used in target speaker voice activity detection (TS-VAD) to calculate the attractor, which can  mitigate the speaker permutation problem and facilitate easier model convergence. In the training process, we propose a teacher-forcing strategy to obtain the enrollment information using the ground-truth label. Furthermore, we propose three heuristic decoding methods to identify the enrollment area for each speaker during the evaluation process. Additionally, we enhance the attractor calculation network LSTM used in the end-to-end encoder-decoder based attractor calculation (EEND-EDA) system by incorporating an attention-based model. By utilizing such an attention-based attractor decoder, our proposed AED-EEND system outperforms both the EEND-EDA and TS-VAD systems with only 0.5s of enrollment data.


\end{abstract}

\section{Introduction}
The speaker diarization task is defined to address the  ``Who spoke when?'' problem. Traditional speaker diarization systems \cite{shum2013unsupervised,sell2014speaker} usually use a stage-wise paradigm: 1) A speaker embedding extractor is adopted to get the speaker embedding for each segment. 2) A clustering algorithm \cite{sell2018diarization,wang2018speaker,lin2019lstm} merges segments based on the speaker embedding similarity and assigns the speaker label to each segment. 3) Optionally, some compensation algorithm like Variational-Bayesian refinement \cite{sell2015diarization,sell2018diarization,diez2018speaker} is used to calibrate the clustering results. However, the stage-wise system can not deal with the overlap problem due to the frame-wise one-to-one assignment of the clustering algorithm.

To address this issue, many end-to-end methods have been proposed in recent years. For example, EEND \cite{fujita2019end_lstm, fujita2019end_sa} models the diarization task as a multi-class classification problem, allowing frames to be classified into multiple speakers. However, a limitation of EEND is its fixed number of output heads, which makes it difficult to handle varying numbers of speakers. To address this limitation, Horiguchi et al. \cite{horiguchi2020end} proposed EEND-EDA, which uses an LSTM encoder-decoder to predict the attractor for each speaker. The number of attractors can be flexible, allowing the system to generalize to sessions where the number of speakers varies. Fujita et al. \cite{fujita2020neural} and Takashima et al. \cite{takashima2021end} have also proposed methods that sequentially output diarization results for each speaker using a chain rule. However, all of these methods suffer from the speaker permutation problem \cite{hershey2016deep}, making system training challenging.

The target-speaker voice activity detection (TS-VAD) \cite{medennikov2020target,medennikov2020stc} system has attracted significant attention due to its outstanding performance in speaker diarization challenges \cite{medennikov2020stc,wang2022dku}. This has led to the development of various TS-VAD variants  \cite{cheng2022multi,wang2022target,cheng2022target,jiang2023target} by different researchers. In the TS-VAD system, a separate diarization system first predicts the single-speaker speaking area for each speaker. Subsequently, the speaker embeddings extracted for each speaker are concatenated with the acoustic features and fed into the diarization system. Because the TS-VAD system uses pre-acquired speaker embeddings, it does not suffer from the speaker permutation problem.

In this paper, we propose an advanced attention-based encoder-decoder network for end-to-end neural speaker diarization (AED-EEND) by integrating the target speaker enrollment information used in the TS-VAD system into the original EEND-EDA system. Our system overcomes the limitations of both TS-VAD and EEND-EDA while leveraging their advantages. Unlike the TS-VAD system, which requires enrollment information from an external diarization system and speaker embedding extractor, our system obtains the enrollment information directly from the AED-EEND system, thereby making our system fully end-to-end. Additionally, we replace the LSTM-based model used in EEND-EDA to generate attractors with an attention-based decoder, which has shown to be more effective.
More importantly, our AED-EEND system only uses an attention-based encoder-decoder network, which is much simpler than the EEND-EDA and TS-VAD systems.

Besides, we propose a teacher forcing strategy for the training process that uses ground-truth labels to acquire enrollment information. For evaluation, we propose three heuristic decoding methods to output the diarization results for each speaker iteratively. Our proposed attention-based attractor decoder is particularly noteworthy as it produces robust results with only 0.5 seconds of enrollment speech, which significantly outperforms both the EEND-EDA and TS-VAD systems on the \textsc{CallHome} evaluation set.


\section{Method}
\label{sec:method}

\subsection{Attention-based Encoder-Decoder Network for EEND}
\label{ssec:aed-eend_model}

\begin{figure}[ht!]
  \centering
  \includegraphics[width=0.46\textwidth]{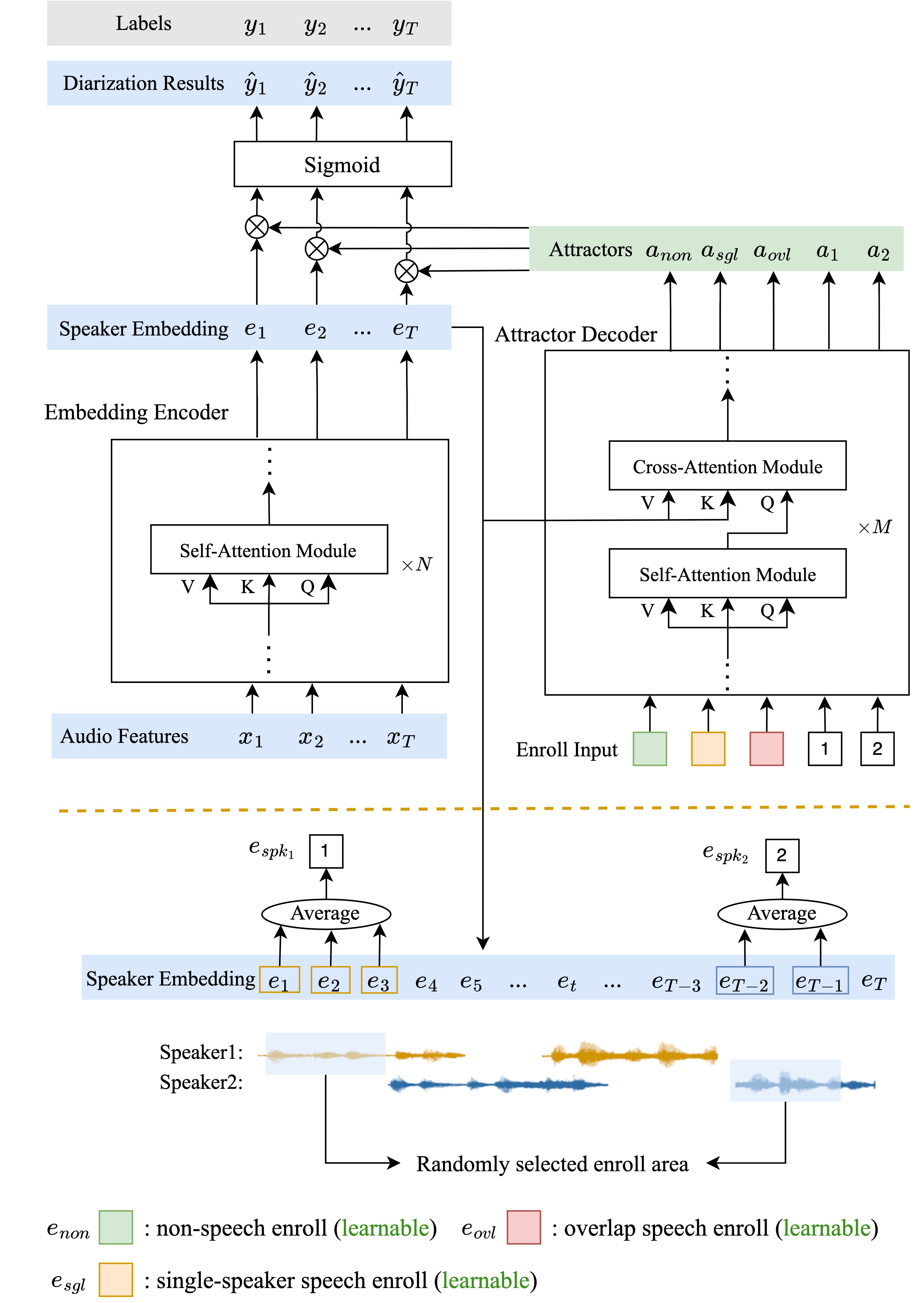}
  \caption{\textbf{AED-EEND system architecture when the speaker number is two.} }
  \label{fig:system_architecture}
\end{figure}

In this section, we introduce the architecture of our proposed AED-EEND system. The overall architecture of our system is shown in the upper part of Figure \ref{fig:system_architecture}. The AED-EEND system contains two parts: an embedding encoder and an attractor decoder. The encoder-decoder architecture is very similar to the original transformer model proposed in \cite{vaswani2017attention} except that we neglect the positional encoding in our system. The embedding encoder accepts the
acoustic feature sequence $X = [\mathbf{x}_1, \mathbf{x}_2, ..., \mathbf{x}_T] \in \mathbb{R}^{T \times F}$ as input, and outputs the frame level speaker embedding sequence $E = [\mathbf{e}_1, \mathbf{e}_2, ..., \mathbf{e}_T]  \in \mathbb{R}^{T \times D}$. The attractor decoder takes the enrollment sequence $E_\text{enroll} = [\mathbf{e}_\text{non}, \mathbf{e}_\text{sgl}, \mathbf{e}_\text{ovl}, \mathbf{e}_{\text{spk}_1}, ..., \mathbf{e}_{\text{spk}_S}]  \in \mathbb{R}^{(S+3) \times D}$ as input, where $\mathbf{e}_\text{non}$, $\mathbf{e}_\text{sgl}$ and $\mathbf{e}_\text{ovl}$ correspond to the enrollment embeddings for non-speech, single-speaker speaking speech and overlap speech, respectively, and $\mathbf{e}_{\text{spk}_i}$ corresponds to the enrollment embedding for $i$-th speaker in the recording. The enrollment embedding sequence $E_\text{enroll}$ is fed into the attractor decoder to get the corresponding attractor sequence $A = [\mathbf{a}_\text{non}, \mathbf{a}_\text{sgl}, \mathbf{a}_\text{ovl}, \mathbf{a}_{\text{spk}_1}, ..., \mathbf{a}_{\text{spk}_S}]  \in \mathbb{R}^{(S+3) \times D}$.

In contrast to the embedding encoder that solely incorporates the self-attention (SA) module, the attractor decoder is equipped with both the self-attention (SA) module and the cross-attention (CA) module. The SA module in the attractor decoder enables the enrollment inputs to attend to one another, whereas the CA module facilitates incorporating additional relevant information from the embedding sequence into the enrollment embedding. Specifically, the output from the SA module serves as the query, while the speaker embedding sequence output from the embedding encoder is treated as the key and value. This mechanism is expected to enhance the quality of the enrollment embedding.

Unlike the EEND-EDA architecture \cite{horiguchi2020end}, which only generates attractors for the speakers present in the recording, our proposed AED-EEND system also outputs attractors for three distinct speech types in the recording. These speech types include the single-speaker, speaker-overlap, and non-speech regions. The resulting attractors are then used to compute frame-level posterior probabilities for each speaker and speech type by performing a dot product operation with the corresponding embedding sequence.

\begin{equation}
\hat{Y}=\sigma\left(A E^{\top}\right) \in(0,1)^{(S+3) \times T}
\end{equation}
where $\sigma(\cdot)$ is the element-wise sigmoid function. Here, we denote the posterior probabilities at each frame as $\hat{\mathbf{y}}_{t} =[\hat{y}_t^{\text{non}}, \hat{y}_t^{\text{sgl}}, \hat{y}_t^{\text{ovl}}, \hat{y}_t^{1}, ..., \hat{y}_t^{S}] \in(0,1)^{(S+3)}$, and denote the ground truth label at each frame for speaker and speech activity as $\mathbf{y}_{t} =[y_t^{\text{non}}, y_t^{\text{sgl}}, y_t^{\text{ovl}}, y_t^{1}, ..., y_t^{S}] \in\{0,1\}^{(S+3)}$. The label $1$ corresponds to the specific speaker or specific kind of speech existing in the corresponding frame, and vice versa. We calculate the loss by averaging the cross entropy between the prediction and ground truth for each speaker or speech type at each frame:
\begin{equation}
\mathcal{L}=\frac{1}{T(S+3)}\sum_{t=1}^T\sum_{s\in \mathbb{S}} \left[ -y_t^s \log \hat{y}_t^s-\left(1-y_t^s\right) \log \left(1-\hat{y}_t^s\right)\right]
\end{equation}
where $\mathbb{S} = \{\text{non}, \text{sgl}, \text{ovl}, 1, .., S\}$.




\subsection{System Training with Teacher Forcing Enrollment}
\label{ssec:system_training}

Similar to the TS-VAD method \cite{medennikov2020target}, our proposed AED-EEND system also requires enrollment embedding for each speaker, but we obtain it directly from the embedding sequence $E$, eliminating the need for a pre-trained speaker embedding extraction system. As depicted in the lower part of Figure \ref{fig:system_architecture}, given the single-speaker speaking area of each speaker in the recording, we can obtain the corresponding embeddings and average them to obtain the enrollment embedding for each speaker.

In the training of automatic speech recognition (ASR) systems, the teacher forcing strategy \cite{woodward2020confidence} is a widely adopted technique that leverages ground-truth input sequences to predict subsequent outputs during the training process. This strategy is known to improve training stability and efficiency. In the training process of our proposed AED-EEND, we similarly employ the teacher-forcing strategy. Specifically, we obtain the single-speaker speaking area for each speaker from the ground-truth label and randomly select a consecutive area with an enrollment length $L_{enroll}$ as the enrollment area.

In contrast to speaker enrollment embeddings, which are obtained from the embedding sequence $E$, we directly set the enrollment embeddings for non-speech, single-speaker speech, and overlapping speech as three learnable embeddings.

\subsection{System Inference by Iterative Decoding}
\label{ssec:decoding}
\input{algorithm/decoding}

In this section, we propose an iterative decoding method for AED-EEND in the evaluation process, and the pipeline is shown in algorithm \ref{algo:decoding}. The first step is to predict the results for single-speaker, overlap, and non-speech with learnable embeddings. Then, the prediction for each speaker is obtained iteratively. At each iteration, we first find the un-predicted single-speaker speaking area $I$ and then split the area into some time-continuous segments. If the longest segment is shorter than a pre-defined threshold $L_{\text{stop}}$, we will stop the decoding. To get the prediction for each speaker in one iteration, we propose three heuristic strategies to enlarge the probability that there is only one speaker in the enrollment area:
\begin{itemize}
    \item \textbf{Init-Decode}: Use the initial $L_{\text{enroll}} $ frames of the first time-continuous segments in $I$ as the enrollment area.
    \item \textbf{Rand-Decode}: Randomly select a time-continuous segment $I'$ from $I$, and then randomly select $L_{\text{enroll}} $ length consecutive frames from $I'$ as the enrollment area.
    \item \textbf{SC-Decode}: Cluster the embeddings in $I$ using spectral clustering algorithm \cite{wang2018speaker,lin2019lstm}, and then randomly select $L_{\text{enroll}} $ length consecutive frames from the biggest cluster as the enrollment area. 
\end{itemize}

Besides, for comparison, we also follow the strategy in the training process to get the enrollment area and denote this strategy as \textbf{GT-Decode}.

\section{Experimental Setup}
To ensure comparability with previous studies, we adopted most of the configurations used in EEND-EDA \cite{horiguchi2020end} except for the model architecture. In \cite{horiguchi2020end}, the authors utilized an LSTM encoder-decoder to output the attractors. As outlined in section \ref{ssec:aed-eend_model}, we replaced the LSTM with a four-layer transformer model. The attention unit number is set to 256 in our experiment. Besides, our experiment also utilized the 345-dimensional acoustic features in \cite{horiguchi2020end} as input to the embedding encoder.

Regarding training data, we followed the pipeline in \cite{horiguchi2020end} to simulate 1-, 2-, 3-, and 4-speaker datasets. However, unlike \cite{horiguchi2020end}, we only created the simulation evaluation set with a similar overlap ratio to the training set. We used the \textsc{CallHome} \cite{callhome} dataset to evaluate our system on real recordings and split the dataset into two parts - part1 for adaptation and part2 for evaluation, in accordance with the Kaldi recipe \footnote{\url{https://github.com/kaldi-asr/kaldi/tree/master/egs/callhome_diarization/v2}}.

Following \cite{horiguchi2020end}, we evaluated our systems under two conditions: the fixed number of speakers condition and the flexible number of speakers condition. For the fixed number of speakers condition, we trained the systems on 2-speaker or 3-speaker datasets for 100 epochs. To evaluate the systems on real recordings, we further fine-tuned the models on the corresponding number of speakers adaptation set from \textsc{CallHome} for another 100 epochs. For the flexible number of speakers condition, we first fine-tuned the pre-trained model on the 2-speaker simulation set on the concatenation of 1-, 2-, 3-, and 4-speaker datasets for another 25 epochs. To evaluate the model on \textsc{CallHome}, we further fine-tuned the model on the entire \textsc{CallHome} adaptation set for another 100 epochs.

During our training process, we set the enrollment length $L_{enroll}$, as introduced in section \ref{sec:method}, to 10-30 frames, corresponding to 1s-3s. During the evaluation process, we set the enrollment length $L_{enroll}$ to 5 frames (0.5s) and the stop decoding length $L_{stop}$ to 10 frames (1s). Additionally, we randomly set some enrollment speaker embeddings to zero vectors during the training process to ensure the model works correctly even when not all the enrollment embeddings are available.

We evaluated each system's performance using the diarization error rate (DER) with a 0.25s collar tolerance to mitigate potential labeling errors.

\section{Results and Analysis}

\subsection{Comparison Among Different Decoding Methods}
\input{tables/decode_method_compare.tex}
This section presents the evaluation results of different decoding methods proposed in section \ref{ssec:decoding}. The results are summarized in Table \ref{table:decode_method_compare}. In the experiments, we either use the oracle speaker number or estimate the speaker number using the stop decoding criterion proposed in section \ref{ssec:decoding}. Surprisingly, the performance gap between estimated speaker number and oracle speaker number is not significant, which demonstrates the effectiveness of the stop decoding criterion in accurately predicting the speaker number. Furthermore, the SC-Decode method exhibits the best performance on most evaluation conditions, even surpassing the GT-Decode method. Therefore, we adopt the SC-Decode method in the subsequent experiments.

\subsection{Results on Fixed Number of Speakers}

\input{tables/fixed_spk_num_res.tex}
This section evaluates our system when the speaker number is fixed and known in advance. The corresponding results are shown in Table \ref{table:fixed_spk_num_res}. Apart from the 2-spk simulation dataset evaluation, our proposed method outperforms all the other methods. It is worth noting that even though our proposed AED-EEND method performs worse than the EEND-EDA on the 2-spk simulation dataset, our method outperforms the EEND-EDA on the 2-spk \textsc{CallHome} dataset, which demonstrates the better generalizability of our proposed method.

\subsection{Results on Flexible Number of Speakers}
\input{tables/flexible_spk_num_simu_res.tex}

\subsubsection{Results on simulation dataset}

In this section, we evaluate our systems when the speaker number is flexible. We first do the experiment on the simulation dataset and show the result in Table \ref{table:flexible_spk_num_simu_res}. From the results, we find our method outperforms both the stage-wise x-vector clustering method and the EEND-EDA method on any number of speakers condition. 

\input{tables/flexible_spk_num_callhome_res.tex}
\subsubsection{Results on \textsc{CallHome} dataset}

Then, we evaluate our system on the \textsc{CallHome} dataset and list the results in Table \ref{table:flexible_spk_num_callhome_res}. The overall performance of our proposed method is better than both x-vector clustering and EEND-EDA, and this improvement mainly comes from the small speaker number condition. 
Similar to the conclusion on the simulation dataset, whether or not to use an oracle speaker number in the decoding process has little effect on the results of \textsc{CallHome} dataset. Besides, similar to the results of EEND-EDA, our system performs worse than the x-vector clustering in the 5-speaker and 6-speaker conditions. In the follow-up work \cite{horiguchi2022encoder} of EEND-EDA, the authors improve these results by simulating the data with more speakers in each recording. We plan to try this approach in our future work.

\subsection{Analysis of Non-Speech, Single-speaker Speech and Overlap Speech Prediction}
\input{tables/silence_single-spk_overlap_res-new}

As mentioned in section \ref{ssec:decoding}, we first predict the non-speech, single-speaker speech, and overlap speech area for an input utterance and then use the single-speaker speech area as the prior knowledge to do the iterative decoding. This section presents an evaluation of our proposed system on these three types of predictions, as shown in Table \ref{table:silence_spk_overlap}. Here, we will treat each type of prediction independently, without considering the potential conflicts between different types of predictions. The simulation dataset yields accurate area predictions for all three types of speech, with false alarm and miss rates below 10\%. However, the system performs worse on some evaluation metrics for the \textsc{CallHome} dataset. We believe that the reason for this outcome is the significant mismatch in single-speaker and overlap proportions between the simulation dataset and the real \textsc{CallHome} dataset. The statistics in \cite{horiguchi2022encoder} indicate that in the simulation dataset, the proportions of different speech types are relatively balanced. However, in the real \textsc{CallHome} dataset, single-speaker speech accounts for the majority of the audio, with low proportions of both overlap and single-speaker speech.


\section{Conclusion}
In this paper, we propose the AED-EEND system, which effectively integrates enrollment information into the EEND-EDA system. To enhance the model's training, we introduce a teacher forcing technique that leverages ground-truth labels to obtain enrollment embeddings during training. We present three heuristic decoding approaches to identify the enrollment information for each speaker during the evaluation process. Furthermore, we replace the LSTM architecture for attractor calculation in EEND-EDA with an attention-based model. Our approach achieves remarkable performance with very short enrollment data. AED-EEND introduces a novel paradigm for diarization tasks. However, due to space limitations, our evaluation on real datasets was limited. In future research, we will conduct a more thorough analysis of our proposed AED-EEND system and evaluate it in a broader range of real-world datasets.

\section{Acknowledgements}
This work was supported in part by China STI 2030-Major Projects under Grant No. 2021ZD0201500, in part by China NSFC projects under Grants 62122050 and 62071288, and in part by Shanghai Municipal Science and Technology Major Project under Grant 2021SHZDZX0102. The author Zhengyang Chen is supported by Wu Wen Jun Honorary Doctoral Scholarship, AI Institute, Shanghai Jiao Tong University.

\bibliographystyle{IEEEtran}
\bibliography{mybib}

\end{document}

%% file: algorithm/decoding.tex
\begin{algorithm}[!ht]
\footnotesize

\KwData{
$I = [t_1, t_2, ...]$: frame indexes list for embedding in $E$ \\
\hspace{20pt} $C = [I_1, I_2, ..., I_K]$: $K$ time-continuous segments\\
\hspace{24pt} from $I$, and the indexes in $I_{k}$ are sorted in order\\




}

 \tcp{get the active frame indexes list for three kinds of speeches}
 $I_{\text{non}}, I_{\text{sgl}}, I_{\text{ovl}}  = \text{AED-EEND}(\mathbf{e}_\text{non}, \mathbf{e}_\text{sgl}, \mathbf{e}_\text{ovl})$ \\
 \tcp{active frame indexes list for speakers; enroll embedding set}
 $I_{\text{spk}}=[]; \mathcal{E}=\{\}$

 \While{True}{
        \tcp{get the un-predicted single-speaker area}
        $I = I_{\text{sgl}} - I_{\text{sgl}} \cap I_{\text{spk}}$ \\

        \tcp{get segments with length $\geq L_{\text{Enroll}}$}
        $C' = [I'_1, I'_2, ..., I'_{K'}] = \text{filter\_segs}(C, L_{\text{Enroll}})$ \\
        \tcp{get the segment with the longest length}
            $I_{\text{longest}} = \text{get\_longest\_seg}(C)$ \\
        
        $C'\text{.add}(I_{\text{longest}})$ \\
        $L_{\text{Enroll\_tmp}} = \text{min}(\text{length}(I_{\text{longest}}), L_{\text{Enroll}})$ \\
        \If{$L_{\text{\text{stop}}} > \text{length}(I_{\text{longest}})$}{break;} 

        \Case{\textbf{Init-Decode}}{
            
            $I_{\text{enroll}} = [t_1, t_2, ..., t_{L_{\text{Enroll\_tmp}}}] \in I'_{1}$
            
        }
        \Case{\textbf{Rand-Decode}}{
            \tcp{randomly select a segment from $C'$}
            $I'_k = \text{random\_select\_seg}(C')$ \\
            \tcp{randomly select a consecutive sub-segment with length $L_{\text{Enroll\_tmp}}$}
            $I_{\text{enroll}} = \text{random\_select\_sub-seg}(I'_k, L_{\text{Enroll\_tmp}})$ \\
        
        }
        \Case{\textbf{SC-Decode}}{
            \tcp{The index in $I$ is clustered based on corresponding embedding}
            $[I^{\text{cls}}_1, I^{\text{cls}}_2, ...] = \text{spectral\_cluster}(I)$ \\
            $I^{\text{cls}}_k = \text{get\_longest\_seg}([I^{\text{cls}}_1, I^{\text{cls}}_2, ...])$ \\
            $I_{\text{enroll}} = \text{random\_select\_sub-seg}(I^{\text{cls}}_k, L_{\text{Enroll\_tmp}})$ \\
            
        }
        \tcp{average the embeddings with indexes in $I_{\text{enroll}}$}
        $\mathbf{e} = \text{average}(E_{I_{\text{enroll}}})$\\
        $\mathcal{E}\text{.add}(\mathbf{e})$\\ 
        $I_{\text{spk}} = (\text{AED-EEND}(\mathcal{E}))$

}
\KwOut{$I_{\text{spk}}$}

\caption{Iterative Decoding Pipeline}
\label{algo:decoding}
\end{algorithm}

%% file: tables/decode_method_compare.tex
\begin{table}[ht!]
\vspace{5pt}
\footnotesize
\centering
\caption{\textbf{DER (\%) results for different decoding methods on 2-speaker evaluation set.}
}
\begin{adjustbox}{width=.47\textwidth,center}
\begin{tabular}{l cccc}
\toprule
\multirow{2}{*}{Dataset} & \multicolumn{4}{c}{Decoding Methods} \\
\cmidrule(r){2-5} 
 & GT-Decode & Init-Decode & Rand-Decode & SC-Decode \\
 \hline
Simulation  & & & & \\
\hspace{3pt} Estimated \#spk &  - & 3.40 & 3.50 & \textbf{3.25}  \\
\hspace{3pt} Oracle \#spk &  3.22 & 3.14 & \textbf{3.13} & 3.14\\
\textsc{CallHome}   & & & & \\
\hspace{3pt} Estimated \#spk &  - & 10.7 & 8.58 & \textbf{8.34} \\
\hspace{3pt} Oracle \#spk &  8.03 & 10.9 & 8.32 & \textbf{7.75} \\

\bottomrule
\end{tabular}
\label{table:decode_method_compare}
\end{adjustbox}
\end{table}

%% file: tables/fixed_spk_num_res.tex
\begin{table}[ht!]
\footnotesize
\centering
\caption{\textbf{DER (\%) results on the fixed number of speakers condition.}
}
\begin{adjustbox}{width=.47\textwidth,center}
\begin{tabular}{lcccc}
\toprule
\multirow{2}{*}{Method} & \multicolumn{2}{c}{Simulation DER(\%)}  & \multicolumn{2}{c}{\textsc{CallHome} DER(\%)}\\
\cmidrule(r){2-3} \cmidrule(r){4-5} 
 & 2-spk & 3-spk & 2-spk & 3-spk \\
 \hline
 x-vector clustering \cite{horiguchi2020end}  &  28.77 & 31.78 & 11.53 & 19.01 \\
 BLSTM-EEND \cite{fujita2019end_lstm} & 12.28 & - & 26.03  & - \\
 SA-EEND  \cite{fujita2019end_sa,horiguchi2020end}  & 4.56   & 8.69  & 9.54  & 14.00 \\
 EEND-EDA  \cite{horiguchi2020end} & \textbf{2.69}   & 8.38  & 8.07  & 13.92 \\
 TS-VAD \cite{cheng2022multi} & - & - & 9.51 & - \\
 AED-EEND (ours)                   & 3.14  & \textbf{5.16} & \textbf{7.75}  & \textbf{12.87}  \\

\bottomrule
\end{tabular}
\label{table:fixed_spk_num_res}
\end{adjustbox}
\end{table}

%% file: tables/flexible_spk_num_simu_res.tex
\begin{table}[ht!]
\footnotesize
\centering
\caption{\textbf{DER (\%) results on the simuation dataset with flexible number of speakers.}
}
\begin{adjustbox}{width=.36\textwidth,center}
\begin{tabular}{lcccc}
\toprule
\multirow{2}{*}{Method} & \multicolumn{4}{c}{spk\#}\\
\cmidrule(r){2-5}
 & 1 & 2 & 3 & 4 \\
 \hline
 x-vector clustering \cite{horiguchi2020end} & & & \\
 \hspace{3pt} Estimated \#spk & 37.42 & 7.74 & 11.46 & 22.45 \\
 \hspace{3pt} Oracle \#spk & 1.67 & 28.77 & 31.78 & 35.76 \\

 EEND-EDA \cite{horiguchi2020end} & & & \\
 \hspace{3pt} Estimated \#spk & 0.39 & 4.33 & 8.94 & 13.76 \\
 \hspace{3pt} Oracle \#spk & 0.16 & 4.26 & 8.63 & 13.31 \\
 
 AED-EEND (ours) & & & \\
 \hspace{3pt} Estimated \#spk & \textbf{0.09} & \textbf{2.82} & 6.88 & 12.00 \\
 \hspace{3pt} Oracle \#spk & \textbf{0.09} & \textbf{2.82} & \textbf{6.57}  & \textbf{11.50} \\


\bottomrule
\end{tabular}
\label{table:flexible_spk_num_simu_res}
\end{adjustbox}
\vspace{-10pt}
\end{table}

%% file: tables/flexible_spk_num_callhome_res.tex
\begin{table}[ht!]
\footnotesize
\centering
\caption{\textbf{DER (\%) results on the \textsc{CallHome} dataset with flexible number of speakers.}
}
\begin{adjustbox}{width=.47\textwidth,center}
\begin{tabular}{lcccccc}
\toprule
\multirow{2}{*}{Method} & \multicolumn{6}{c}{spk\#}\\
\cmidrule(r){2-7}
 & 2 & 3 & 4 & 5 & 6 & all\\
 \hline
 x-vector clustering \cite{horiguchi2020end} & & & & &  \\
 \hspace{3pt} Estimated \#spk & 15.45 & 18.01 & 22.68 & \textbf{31.40} & \textbf{34.27} & 19.43 \\
 \hspace{3pt} Oracle \#spk & 8.93 & 19.01 & 24.48 & 32.14 & 34.95 & 18.98 \\

 EEND-EDA \cite{horiguchi2020end} & & & \\
 \hspace{3pt} Estimated \#spk & 8.50 & 13.24 & 21.46 & 33.16 & 40.29 & 15.29 \\
 \hspace{3pt} Oracle \#spk & 8.35 & 13.20 & 21.71 & 33.00 & 41.07 & 15.43 \\


 AED-EEND (ours) & & & \\
 \hspace{3pt} Estimated \#spk & 6.96 & 12.56 & \textbf{18.26} & 34.32 & 44.52 & 14.22 \\
 \hspace{3pt} Oracle \#spk & \textbf{6.79} & \textbf{12.36} & 19.84 & 32.42 & 37.08 & \textbf{14.00} \\

 

\bottomrule
\end{tabular}
\label{table:flexible_spk_num_callhome_res}
\end{adjustbox}
\vspace{-10pt}
\end{table}

%% file: tables/silence_single-spk_overlap_res-new.tex
\begin{table}[ht!]
\vspace{-10pt}
\footnotesize
\centering
\caption{\textbf{AED-EEND result (\%) on the non-speech, single-speaker speech and overlap speech prediction.} 
$\text{FA}$ represents the false alarm rate (\%) or false positive rate (\%). $\text{MISS}$ represents the miss error rate (\%) or false negative rate (\%). All the results are based on our 2-spk systems. 
}
\begin{adjustbox}{width=.47\textwidth,center}
\begin{tabular}{c|cc|cc|cc}
\toprule
 \multirow{2}{*}{Dataset}  & \multicolumn{2}{c}{Non-Speech} & \multicolumn{2}{c}{Single-Speaker} & \multicolumn{2}{c}{Overlap}\\
\cline{2-7}
                    & $\text{FA}$ & $\text{MISS}$ & $\text{FA}$ & $\text{MISS}$ & $\text{FA}$ & $\text{MISS}$ \\
\hline
Simulation & 0.91 & 3.57 & 6.93 & 6.25 & 3.44 & 9.59 \\
  \textsc{CallHome} & 3.20 & 30.68 & 35.61 & 7.62 & 3.51 & 41.07 \\

\bottomrule
\end{tabular}
\label{table:silence_spk_overlap}
\end{adjustbox}
\vspace{-5pt}
\end{table}